\def\beq{\begin{equation}}
\def\eeq{\end{equation}}
\def\beqa{\begin{eqnarray}}
\def\eeqa{\end{eqnarray}}
\def\darr{\mathrel{\hbox{\rlap{\hbox{\lower1.0pt\hbox{$\partial_\mu$}}}\hbox{\raise5.0pt\hbox{$\!\leftrightarrow$}}}\!}}

\def\beq{\begin{equation}}
\def\eeq{\end{equation}}
\def\bea{\begin{eqnarray}}
\def\eea{\end{eqnarray}}
\def\bq{\begin{quote}}
\def\eq{\end{quote}}

\def\bq{\begin{quote}}
\def\eq{\end{quote}}

\documentstyle[pramana,epsf,floats]{ias}
\begin{document}
\mark{{Non-Minimal and Non-Universal Supersymmetry}{S.F.\ King}}
\title{Non-Minimal and Non-Universal Supersymmetry}

\author{S.F.\ King}
\address{Dept. of Physics and Astronomy, University
of Southampton, Southampton SO17 1BJ, U.K.}
\abstract{I motivate and discuss non-minimal and non-universal 
models of supersymmetry and supergravity
consistent with string unification at $10^{16}$ GeV.
}

\maketitle
\section{Introduction}
The motivations for TeV scale supersymmetry (SUSY)
\cite{susy} remain as good as
ever:
\begin{enumerate}
\item TeV scale SUSY cancels the quadratic divergences in the Higgs mass
(hierarchy problem),
\item TeV scale SUSY ensures that the gauge couplings meet at 
$M_{GUT}\sim 10^{16} $ GeV.
\end{enumerate}
The discovery of D-branes allows gravity live in extra dimensions
not felt by the gauge forces, and permits the effective gravitational
coupling $G_NE^2$ to be unified with the other gauge couplings at
$M_{GUT}\sim 10^{16} $ GeV \cite{Ibanez}.
It also opens a Pandora's box of many other higher-dimensional
scenarios for the unification of gravity and gauge interactions,
possibly at scales as low as 1 TeV \cite{Tye}. 
However in many such scenarios
there is no convincing explanation for
the apparent merging of the gauge couplings at $10^{16}$ GeV.
In this talk we shall therefore discuss scenarios consistent with
conventional high energy gauge unification, and ignore 
all large compact dimension scenarios.

Since the vacuum is not manifestly supersymmetric, it must be broken.
Mechanisms for SUSY breaking may be classified according to the
magnitude of the gravitino mass $m_{3/2}$:
\begin{itemize}
\item $m_{3/2}\sim 1$ TeV (gravity mediated \cite{Nilles}) 
\item $m_{3/2}\ll 1$ TeV (gauge mediated \cite{Giudice})
\item $m_{3/2}\gg 1$ TeV (anomaly mediated \cite{RS1})
\end{itemize}
The three mechanism are illustrated schematically in Figs 1-3.
In this talk we shall concentrate mainly on the 
gravity mediated scenario, with the soft masses generated
from an effective supergravity (SUGRA) theory and, according
to our previous discussion, we assume these soft masses
to be generated at a scale of about $10^{16}$ GeV.

\begin{figure}
\epsfxsize=10cm
\epsfysize=3cm
\epsfbox{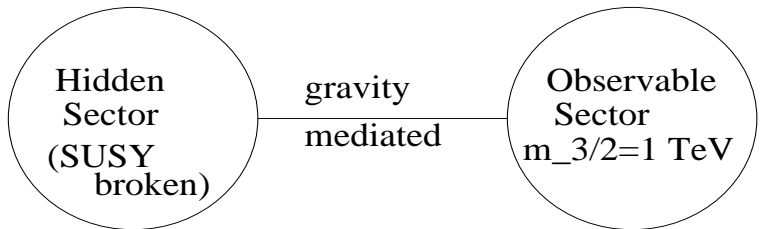}
\caption{{\footnotesize
Gravity mediated SUSY breaking.
There are operators $\sim 1/M_P$ connecting the
hidden sector to the observable sector which communicate
the SUSY breaking.
}}
\end{figure}

\begin{figure}
\epsfxsize=10cm
\epsfysize=7cm
\epsfbox{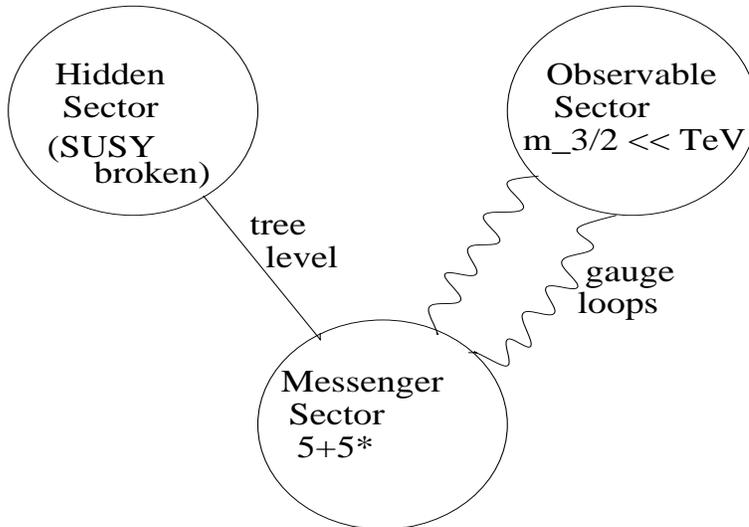}
\caption{{\footnotesize
Gauge mediated SUSY breaking.
There may be operators $\sim 1/M_P$ connecting the
hidden sector to the observable sector but they play no
important role since $m_{3/2}\ll 1 TeV$.
}}
\end{figure}

\begin{figure}
\epsfxsize=10cm
\epsfysize=3cm
\epsfbox{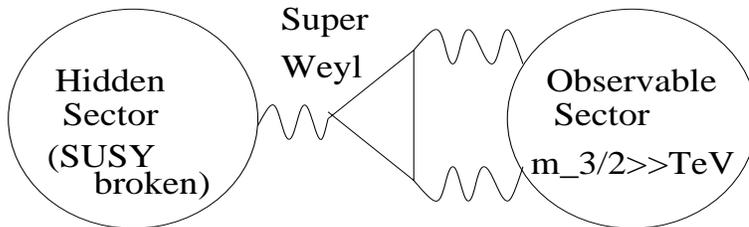}
\caption{{\footnotesize
Anomaly mediated SUSY breaking.
There are {\em no} operators $\sim 1/M_P$ connecting the
hidden sector to the observable sector in this scenario,
which allows the super Weyl anomaly contributions to dominate.
}}
\end{figure}

\section{Elements of a SUGRA theory}
As indicated in Figs. 1-3 the
superpotential and Kahler potential \cite{Nilles} have the form
\bea
W & = & W_{hid}+W_{obs} \nonumber \\
K & = & K_{hid}+K_{obs}
\eea
and the Kahler function 
\beq
G=K/M_{P}^2+\ln |W/M_{P}^3|^2 
\eeq
when inserted into the F-terms of the SUGRA potential
\beq
V_F=M_{P}^4e^G[G^i(G^{-1})^j_iG_j-3]
\eeq
where $G^i=\frac{\delta G}{\delta \phi_i}$,
$G_i=\frac{\delta G}{\delta \phi^{*i}}$,
$(G^{-1})^k_iG^j_k=\delta^j_i$,
leads to a hidden sector SUSY breaking order parameter
$F_i=-M_P^2e^{G/2}(G^{-1})^j_iG_j$
and a gravitino mass
\beq
m_{3/2}^2=\frac{1}{3M_P^2}<K^i_jF_iF^{*j}>=M_P^2e^{<G>}
\eeq
where the last equality assumes $<V_F>=0$.
Terms in the expansion of $V_F$ then lead to soft SUSY
breaking masses in the observable sector depending on the
details of the Kahler potential.
The common feature of the models we consider is that 
the gravitino mass is $m_{3/2} \sim TeV$ which corresponds to 
a SUSY breaking scale in the hidden sector of 
$<F_i> \sim (3\times 10^{10} GeV)^2$.

In minimal SUGRA the Kahler potential is postulated to 
have the form
\beq
K=K_{hid}+\tilde{Q}^{*i}\tilde{Q}^{i}+\ldots
\eeq
where $\tilde{Q}^{i}$ represents one of the squarks or sleptons,
which are thereby assumed to have diagonal metric and minimal
kinetic terms. This results in universal soft scalar masses
\beq
V_{soft}=m_0^2(\tilde{Q}^{*i}\tilde{Q}^{i}+\ldots)
\eeq
where $m_0^2=m_{3/2}^2$,
and universal soft trilinear parameters of order $m_{3/2}$
proportional to the Yukawa couplings in the superpotential.
If the gauge kinetic functions $f_a$ are independent of $a$
then this results in universal gaugino masses $M_{1/2}$ 
of order $m_{3/2}$ at high energies.

In string theory the hidden sector consists of the dilaton $S$
and moduli $T_i$, which get vacuum expectation values (VEVs) 
of order $M_P$.
Although the string mechanism of SUSY breaking
is unclear it may be parametrised in terms of the F-terms of
$S,T_i$ \cite{Ibanez2},
\bea
F^S & = & \sqrt{3}m_{3/2}(S+S^*)\sin \theta e^{-i\gamma_S} 
\nonumber \\
F^i & = & \sqrt{3}m_{3/2}(T_i+T_i^*)\cos \theta \Theta_i e^{-i\gamma_i} 
\eea
where $\sum \Theta_i^2=1$.
The dilaton and moduli couple directly to the MSSM multiplets
in a very complicated non-universal way in the Kahler potential
\beq
K=K_{hid}+\tilde{Q}^{*i}\tilde{Q}^{i}(T_1+T_1^*)^{n_{Q_i}^{T_1}}
(T_2+T_2^*)^{n_{Q_i}^{T_2}}(T_3+T_3^*)^{n_{Q_i}^{T_3}}(S+S^*)^{n_{Q_i}^{S}}
+\ldots
\eeq
where ${n_{Q_i}^{T_1}}$, etc. are the half-integer valued modular
weights, whose $i$ dependence leads to non-universality in the soft scalar
masses,
\beq
m_{Q_i}^2=m_{3/2}^2(1+3\sin^2\theta{n_{Q_i}^{S}}
+3\cos^2\theta{n_{Q_i}^{T_a}.\Theta_a^2)}
\eeq
The trilinears are similarly non-universal. 
The gauge kinetic functions in a D-brane theory \cite{Ibanez2}
involving gauge groups on intersecting 9-branes and 5-branes with
$f_9=S, \ f_{5_a}=T_a$ leads to gaugino masses
\bea
M_9 & = & \sqrt{3}m_{3/2}\sin \theta e^{-i\gamma_S} 
\nonumber \\
M_{5_i} & = & \sqrt{3}m_{3/2}\cos \theta \Theta_i e^{-i\gamma_i} 
\eea
Thus if different standard model gauge factors are on different 
branes the generic expectation is non-universal gaugino masses.
In general the message is clear: from a modern string perspective
non-universality is the natural expectation.

\section{Fine-tuning in SUGRA models}
An issue amongst SUSY phenomenologists at the moment is
the absence of Higgs and superpartners at currently accessible experimental
energies. Is this really a concern?
The question revolves around the issue of how much fine-tuning one
is prepared to tolerate. Although fine-tuning is not a well defined concept,
the general notion of fine-tuning is unavoidable 
since it is the existence of fine-tuning in the
standard model which provides the strongest motivation for low energy 
supersymmetry, and the widespread belief that superpartners should
be found before or at the LHC. Although a precise measure of 
{\em absolute } fine-tuning is impossible,
the idea of {\em relative fine-tuning}
can be helpful in selecting certain models and regions
of parameter space over others. It is useful to compare
different models using a common definition of fine-tuning \cite{BG}
\beq
\Delta_a= abs\left(\frac{a}{M_Z^2}\frac{\partial M_Z^2}{\partial a}\right)
\eeq
where $a$ is an input parameter, and fine-tuning $\Delta^{max}$ is defined
to be the maximum of all the $\Delta_a$.

In a recent paper \cite{BGKK} we compared the fine-tuning of 
several different SUGRA models as listed below: 
\begin{enumerate}

\item Minimal supergravity 
\bea
a_{msugra}\in \{m_0^2, M_{1/2}, A(0), B(0), \mu(0)\}\,,
\label{msugra}
\eea 
where as usual $m_0$, $M_{1/2}$ and $A(0)$ are the universal scalar mass,
gaugino mass and trilinear coupling respectively, $B(0)$ is the
soft breaking bilinear coupling in the Higgs potential and $\mu(0)$ is
the Higgsino mass parameter. 

\item No-scale supergravity with non-universal gaugino masses
\bea
a_{no-scale}\in \{M_1(0), M_2(0), M_3(0), B(0), \mu(0)\}
\label{noscalesugra}
\eea 

\item D-brane model
\bea
a_{D-brane}\in \{m_{3/2}, \theta , \Theta_1, \Theta_2, \Theta_3,
 B(0), \mu(0)\}\,,
\label{Dbrane}
\eea 
where $\theta$ and $\Theta_i$ are the goldstino angles, with
$\Theta_1^2+\Theta_2^2+\Theta_3^2=1$, and $m_{3/2}$
is the gravitino mass. The gaugino masses are given by
\bea
M_1(0)  =  M_3(0) & = & \sqrt{3}m_{3/2}\cos \theta \Theta_1 e^{-i\alpha_1}
\nonumber \,,\\
M_2(0) & = & \sqrt{3} m_{3/2}\cos \theta \Theta_2 e^{-i\alpha_2} \,,
\eea
and there are two types of soft scalar masses
\bea
m_{5152}^2 & = & m_{3/2}^2
[1-\frac{3}{2}(\sin^2 \theta +\cos^2 \theta \Theta_3^2) ]
\nonumber \,,\\
m_{51}^2 & = & m_{3/2}^2[1-3\sin^2 \theta] \,,
\eea

\item Anomaly mediated supersymmetry breaking
\bea
a_{AMSB}\in \{m_{3/2}, m_0^2, B(0), \mu(0)\}
\label{AMSB}
\eea 

\end{enumerate}

Our main results are shown in Figs.4-7, corresponding to SUGRA
models 1-4 above. In all models, 
fine-tuning is reduced as $\tan \beta$ is increased, with
$\tan \beta =10$ preferred over $\tan \beta =2,3$.
Nevertheless,  
the present LEP2 limit on the Higgs and chargino mass
of about 100 GeV and the gluino mass limit of about 250 GeV
implies that $\Delta^{max}$ is of order 10 or higher. 
\begin{figure}
\epsfxsize=10cm
\epsfysize=6cm
\epsfbox{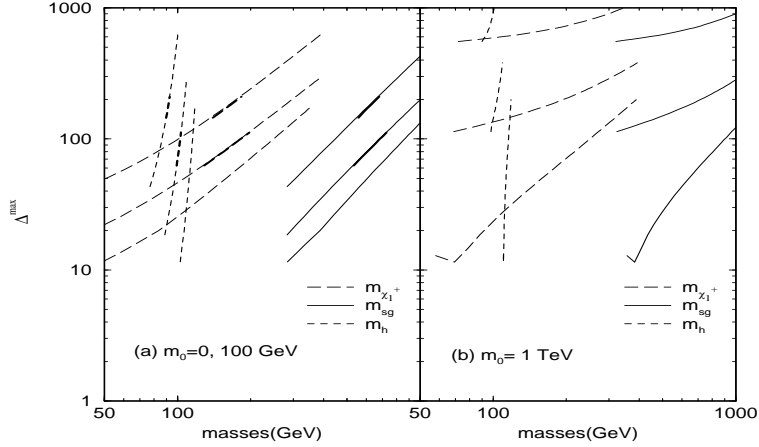}
\caption{ \footnotesize{Results for the minimal SUGRA model.
The maximum sensitivity parameter $\Delta^{max}$ is plotted 
as a function of the lightest CP even Higgs mass (short dashes),
gluino mass (solid line) and lightest chargino (long dashes).
For each particle type, the three sets of curves correspond to 
$\tan\beta$=2, 3, 10, from top left to bottom right, respectively. 
In panel (a) the shorter, thicker lines
correspond to $m_0=0$, while the longer lines are those for $m_0=100$
GeV. In panel (b) the results correspond to $m_0=1000$ GeV.}}
\end{figure}
\begin{figure}
\epsfxsize=10cm
\epsfysize=6cm
\epsfbox{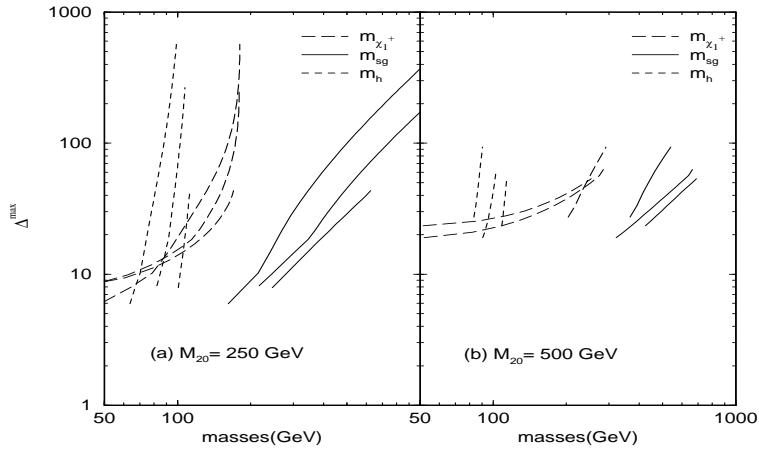} 
\caption{\footnotesize{Results for the 
no-scale with non-universal gaugino masses. 
The maximum sensitivity parameter $\Delta^{max}$ is plotted 
as a function of the lightest CP even Higgs mass (short dashes),
gluino mass (solid line) and lightest chargino (long dashes).
For each particle type, the three sets of curves correspond to 
$\tan\beta$=2, 3, 10, from top left to bottom right, respectively. 
In panel (a) we fix $M_2(0)=250$ GeV, while in panel (b) 
$M_2(0)=500$ GeV.}} 
\end{figure}
\begin{figure}
\epsfxsize=10cm
\epsfysize=6cm
\epsfbox{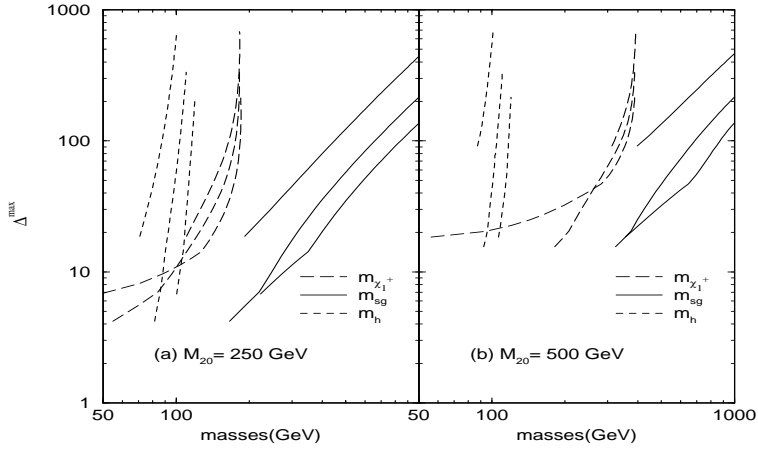}
\caption{{\footnotesize Results for the D-brane model. 
The maximum sensitivity parameter $\Delta^{max}$ is plotted 
as a function of the lightest CP even Higgs mass (short dashes),
gluino mass (solid line) and lightest chargino (long dashes).
For each particle type, the three sets of curves correspond to 
$\tan\beta$=2, 3, 10, from top left to bottom right, respectively.
In panel (a) we fix $M_2(0)=250$ GeV, while in panel (b) 
$M_2(0)=500$ GeV.}}
\end{figure}
\begin{figure}
\epsfxsize=10cm
\epsfysize=6cm
\epsfbox{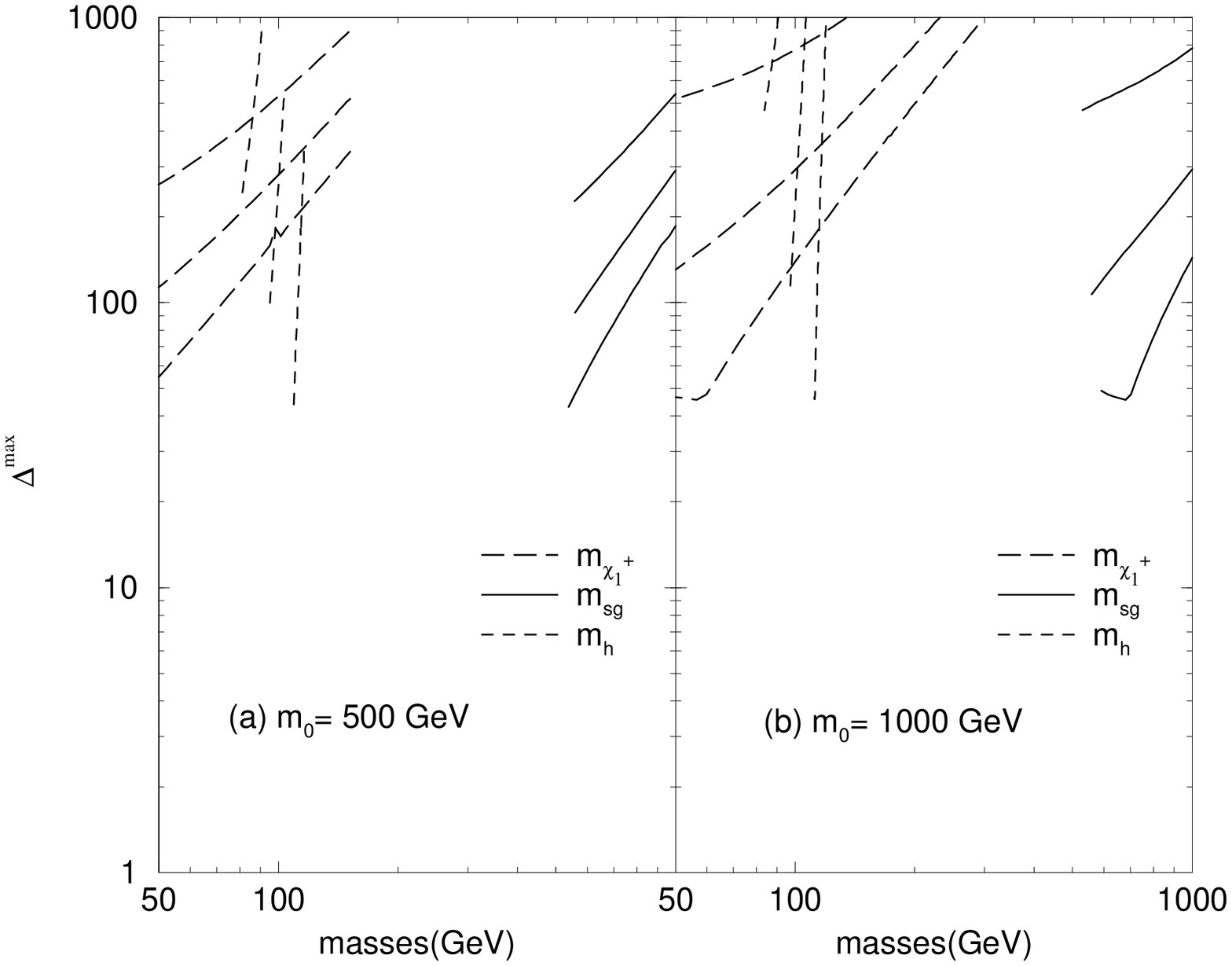}
\caption{{\footnotesize Results for the
anomaly mediated supersymmetry breaking model.
The maximum sensitivity parameter $\Delta^{max}$ is plotted 
as a function of the lightest CP even Higgs mass (short dashes),
gluino mass (solid line) and lightest chargino (long dashes).
For each particle type, the three sets of curves correspond to 
$\tan\beta$=2, 3, 10, from top left to bottom right, respectively.
In panel (a) we fix $m_0=500$ GeV, while in panel (b) 
$m_0=1000$ GeV.}}
\end{figure}
The fine-tuning increases most sharply with the Higgs mass.
The reasons for this are spelled out in the analytic treatment
in ref.\cite{KK}. The starting point for understanding
fine-tuning are the minimal supersymmetric
standard model (MSSM) minimisation conditions
\bea 
\frac{M_Z^2}{2} & = & -\mu^2(t) + \left(
\frac{m_{H_D}^2(t)-m_{H_U}^2(t)\tan^2 \beta}{\tan^2 \beta -1} \right) 
\label{7} \\
\sin 2\beta & = & \frac{2m_3^2(t)}{m_1^2(t)+m_2^2(t)} 
\eea 
where \beq \tan
\beta =v_2/v_1, \ \ M_Z^2=\frac{1}{2} (g'^2+g_2^2)(v_1^2+v_2^2) 
\eeq 
From these conditions the Z mass may be expanded in terms
of high energy input parameters.
For example for $\tan \beta =2.5$ we find \cite{KK},
\bea
\frac{M_Z^2}{2} =  
& & \mbox{} - .87\,\mu^2(0) +
3.6\,{M_3^2(0)}- .12\,{ M_2^2(0)} + .007\,{M_1^2(0)}  \nonumber \\
& & \mbox{} - .71\, {m_{H_U}^2(0)} + .19\,{ m_{H_D}^2(0)}
+ .48\,({ m_Q^2(0)} + \,{m_U^2(0)}) 
\nonumber \\
 & & \mbox{}  - .34\,{A_t(0)}\,{M_3(0)} - .07\,{ A_t(0)}\,{M_2(0)} 
- .01\,{A_t(0)}\,{M_1(0)}
+ .09\,{ A_t^2(0)}\nonumber \\
 & & \mbox{} + .25\,{M_2(0)}\,{M_3(0)}+ .03\,{M_1(0)}\,{M_3(0)} 
+ .007\,{M_1(0)}\,{M_2(0)}
\label{MZ}
\eea
It is clear that $M_3(0)$ dominates the right-hand side, so to reduce
fine-tuning we should make $M_3(0)$ as small as possible.
However $M_3(0)$ cannot be made too small otherwise the gluino mass
becomes too light. More importantly $M_3(0)$ contributes to the
radiative corrections to the Higgs mass. For a given tree-level
Higgs mass the experimental bound implies a lower bound on 
$M_3(0)$ which may be stronger than that coming from the gluino mass
bound. Moreover this bound grows exponentially with the Higgs mass
since the radiatively corrected Higgs mass includes the terms
\beq m_h^2\approx M_Z^2 \cos^2 2 \beta + (34 \ GeV)^2 \ln
\left( \frac{m_{\tilde{t}_1}^2 m_{\tilde{t}_2}^2}{m_t^4}\right) + \ldots 
\label{rc}
\eeq
and the product of stop masses may be expanded as
\bea
{m_{\tilde{t}_1}^2m_{\tilde{t}_2}^2} & \approx &
(170 \ GeV)^4 + 26\,{ M_3^4(0)}+ 1.5\,{ A_t(0)}\,{ M_3^3(0)} \nonumber
\\
& & \mbox{} + \left(2.5\,{ m_Q^2(0)} + 3.5\,{ m_U^2(0)} -2.1\,{m_{H_U}^2(0)}
\right) \,{ M_3^2(0)} \nonumber \\
 & & \mbox{} + (134 \ GeV)^2\,{m_Q^2(0)} + (130 \ GeV)^2\,{m_U^2(0)}
- (108 \ GeV)^2\,{m_{H_U}^2(0)} \nonumber \\
& & \mbox{} + (430 \ GeV)^2\,{ M_3^2(0)}
-(180 \ GeV)^2M_2(0)M_3(0) \nonumber \\
 & & \mbox{}  + (180 \ GeV)^2\,{A_t(0)}\,{M_3(0)} - (59 \ GeV)^2\,{A_t^2(0)}
- (65 \ GeV)^2\,\mu^2 (0) \nonumber \\
 & & \mbox{} + (67 \ GeV)^2\,{A_t(0)}\,\mu (0)
- (210 \ GeV)^2\,{M_3(0)}\,\mu (0) 
\label{det}
\eea
Taken together Eqs.\ref{rc} and \ref{det} show that any
shortfall in the tree-level contribution to the Higgs mass
must be compensated by
exponential increases in stop masses, which in turn involves exponential
increases in $M_3 (0)$, and hence exponential increases in fine-tuning.

The Higgs fine-tuning curves are fairly model independent,
and as the Higgs mass limit rises above 100 GeV come to 
quickly dominate the fine-tuning. We conclude that the 
prospects for the discovery of the Higgs boson at LEP2 are good.
For each model there is a correlation between the
Higgs, chargino and gluino mass, for a given value of
fine-tuning. For example if the Higgs is discovered
at a particular mass value, then the corresponding
chargino and gluino mass for each $\tan \beta$ can be 
read off from Figs.4-7. 

The new general features of the results may then be summarised as follows:

\begin{itemize}
\item The gluino mass curves are
less model dependent than the chargino curves,
and this implies that in all models if the fine-tuning
is not too large then the prospects for the 
discovery of the gluino at the Tevatron are good.

\item The fine-tuning due to the chargino mass is model
dependent. For example in the no-scale model 
with non-universal gaugino masses and the
D-brane scenario the charginos may be relatively heavy compared
to mSUGRA.

\item Some models have less fine-tuning than others.
We may order the models on the basis of fine-tuning from
the lowest fine-tuning to the highest fine-tuning:
D-brane scenario $<$ generalised no-scale SUGRA $<$ mSUGRA $<$ AMSB. 

\item The D-brane model is less fine-tuned partly because the gaugino
masses are non-universal, and partly because there are large regions
where $\Delta_{m_{3/2}}$, 
$\Delta_{\mu(0)}$, and $\Delta_\theta$ are all close to zero
However in these regions the fine tuning is 
dominated by $\Delta_\Theta$, and this leads to an inescapable
fine-tuning constraint on the Higgs and gluino mass.

\end{itemize}

\section{Can Supersymmetric Soft Phases Be the Source of all CP Violation?}

A further motivation for non-universal SUSY comes from 
the possibility that all of CP violation arises from the phases
present on soft SUSY masses. With universal soft masses this is
impossible, but with non-universal soft masses it has recently 
been shown that even if the CKM phase is zero, it is possible
to account for $\epsilon$ and $\epsilon'$ via gluino exchange
diagrams involving left-handed down squarks mixing with
right-handed strange squarks via a complex soft mass term \cite{MM}.
We have shown \cite{Betal} that CP violation
in the B sector may also arise from chargino-stop box diagrams
with a complex stop mixing mass. A particular structure is required
to avoid excessive contributions from similar diagrams to $\epsilon$.
An important feature is that only {\em flavour independent}
phases (for example phases on the non-universal gaugino masses)
are required. The model predicts $\sin 2\beta = -\sin 2\alpha$
which implies that 
$B\rightarrow \psi K_S$ and $B\rightarrow \pi^+\pi^-$
are related. Further details are summarised in the table.
\begin{center}
\vspace{.5in}
\begin{tabular}{||c||c||c||}
\hline
\hline
Observable & Dominant Contribution & Flavor Content\\ 
\hline
nEDM& $\tilde{g}$, $\tilde{\chi}^{+}$, $\tilde{\chi}^{0}$&
$(\delta_{dd})_{LR}$, $\sim \tilde{K}_{ud}\tilde{K}^{*}_{ud}$\\
$\epsilon$& $\tilde{g}$&
$(\delta_{ds})_{LR}$\\
$\epsilon'$& $\tilde{g}$&                                      
$(\delta_{ds})_{LR}$\\
$\Delta m_K$& SM & 
SM\\
$K_L\rightarrow \pi \nu \bar{\nu}$&SM, $\tilde{g}$
& $(\delta_{ds})_{LR}$\\
$\Delta m_{B_d}$& 
$\tilde{\chi}^{+}$&
$|\tilde{K}_{tb}\tilde{K}^{*}_{td}|$\\
$\Delta m_{B_s}$& SM, $\tilde{\chi}^{+}$&
$|\tilde{K}_{tb}\tilde{K}^{*}_{ts}|$\\
$\sin 2\beta$&$\tilde{\chi}^{+}$&$\tilde{K}_{tb}\tilde{K}^{*}_{td}$\\
$\sin 2\alpha$&$\tilde{\chi}^{+}$&$\tilde{K}_{tb}\tilde{K}^{*}_{td}$\\
$\sin 2\gamma$&$\tilde{\chi}^{+}$&$\tilde{K}_{tb}\tilde{K}^{*}_{ts}$\\
${\cal A}_{CP}(b\rightarrow s \gamma)$&$\tilde{\chi}^{+}$
&$\sim \tilde{K}_{tb}\tilde{K}^{*}_{ts}$\\
$\Delta m_{D}$& $\tilde{g}$& $\sim
|\tilde{K}_{tc}\tilde{K}^*_{tu}|$\\
$n_{B}/n_{\gamma}$& $\tilde{\chi}^{+}$, $\tilde{\chi}^{0}$,
$\tilde{t}_R$& -- \\
\hline
\hline
\end{tabular}
\end{center}

\section{Conclusion}
We have concentrated on conventional
theories in which gauge unification occurs
at a high energy scale $10^{16}$ GeV, and the possibility that
full string unification, including gravity, occurs at this scale.
We have further concentrated on the conventional scenario for 
SUSY breaking, namely gravity mediated SUSY breaking.
In this framework, non-universal soft SUSY masses are the
natural expectation, and the absence of FCNC's remains
an important challenge for SUSY. However there are
phenomenological advantages to having non-universal soft masses
which we have highlighted, namely the reductions in
fine-tuning which arise from having $M_3(0)<M_2(0)$,
and also the resultant lowering of the prediction
for $\alpha_s(M_Z)$ in this case.
Also the possibility that all of CP violation could
arise from flavour-independent phases of soft SUSY masses
is interesting.

The particle mass most sensitive to fine-tuning
is the lightest Higgs boson mass, with fine-tuning growing 
exponentially with the Higgs mass. In this respect it is worth
remembering that we have been assuming the MSSM
based on two Higgs doublets with 
a mass term $\mu H_1 H_2$. In the next-to-minimal supersymmetric
standard model (NMSSM) \cite{NMSSM} the mass term
$\mu$ is replaced by the vacuum expectation value (VEV)
of a Higgs singlet $N$, $\mu H_1 H_2 \rightarrow \lambda N H_1 H_2$.
In the NMSSM the lightest physical CP even Higgs boson
may be heavier than in the MSSM.
The NMSSM also solves the $\mu$-problem and opens the
electroweak baryogenesis window which is tightly constrained
in the MSSM by LEP bounds.

\begin{center}
Acknowledgements
\end{center}
I wish to thank all the organisers, and participants
for making this workshop stimulating and enjoyable.

\end{document}